\documentclass[9pt,twocolumn,twoside]{osajnl}

\journal{josab} 

\setboolean{shortarticle}{false} 

\title{Broadband gain induced Raman comb formation in a silica microresonator}

\author[ ]{Ryo~Suzuki}
\author[ ]{Akihiro~Kubota}
\author[ ]{Atsuhiro~Hori}
\author[ ]{Shun~Fujii}
\author[*]{Takasumi~Tanabe}

\affil[ ]{Department of Electronics and Electrical Engineering, Faculty of Science and Technology, Keio University, 3-14-1, Hiyoshi, Kohoku-ku, Yokohama 223-8522, Japan}

\affil[*]{Corresponding author: takasumi@elec.keio.ac.jp}

\ociscodes{(140.3550) Lasers, Raman; (140.3945) Microcavities; (190.5650) Raman effect.}

\begin{abstract}
A high-Q silica whispering-gallery mode microresonator is an attractive platform on which to demonstrate a broad and phase-locked Raman comb in various wavelength regimes. Raman combs can be used for applications such as compact pulse laser sources, sensors, optical clocks, and coherence tomography. However, the formation dynamics of a Raman comb has not been well exploited. Here we study the dynamics of the Raman comb formation in silica rod microresonators, which have cavity free spectral ranges in microwave rates. We generated a Raman comb with a smooth spectral envelope and also observed the transition between two Raman combs located at different center wavelengths. The transition behavior was obtained when we changed the pump detuning and the coupling strength between the microresonator and fiber. We also explain these phenomena by using a simple model based on coupled mode equations.
\end{abstract}

\setboolean{displaycopyright}{true}

\begin{document}

\maketitle

\section{Introduction} \label{sec:1}
Stimulated Raman scattering (SRS) is an optical nonlinear process that can generate red-shifted light from pump light via the interaction between light and molecular vibrations. The wavelength shift is determined by the vibration mode frequencies of the host material. Although SRS can provide lasing action and optical amplification, it usually requires very high intense pumping. Therefore, previous work relied on pulse excitation for Raman lasing \cite{Stolen_APL_1977,Stolen_JOSAB_1984,Boyraz_OE_2004}. On the other hand, optical microresonators with a high quality factor (Q) and a small mode volume enable us to demonstrate low threshold lasing with continuous wave (CW) pumping. Indeed, microresonator-based Raman lasing with a CW pump has been reported using silica \cite{Spillane_N_2002,Kippenberg_IEEEJSTQE_2004,Ozdemir_PNAS_2014,Ooka_APE_2015,Kato_OE_2017,Yang_NP_2017}, silicon \cite{Rong_NP_2007,Rong_NP_2008}, fluoride materials \cite{Grudinin_OL_2007,Liang_PRL_2010,Lin_OL_2016}, and even other materials \cite{Vanier_OE_2014,Li_OL_2013,Latawiec_O_2015,Liu_O_2017}.\par
Recently, multi-mode Raman lasing, which is often referred as a Raman comb, has been reported \cite{Liang_PRL_2010,Lin_OL_2016,Yang_NP_2017}. A phase-locked Raman comb was demonstrated using a crystalline microresonator in a weak normal dispersion regime \cite{Liang_PRL_2010,Lin_OL_2016}. Even a soliton Raman comb has been demonstrated in a silica disk microresonator in an anomalous dispersion regime, by transferring a four-wave mixing (FWM) comb to a long wavelength regime via Raman processes \cite{Yang_NP_2017}. Although microresonator-based FWM combs usually require an anomalous dispersion, Raman combs have an advantage that they can be generated with a normal dispersion because SRS occurs regardless of the dispersion. Phase-locked Raman combs can be used for such applications as compact pulse laser sources, sensors, optical clocks, and coherence tomography.\par
Despite these various potential applications, the formation dynamics of a Raman comb has not been well understood. Since silica exhibits broadband Raman gain consisting of many molecular vibration modes \cite{Agrawal_2007,Hollenbeck_JOSAB_2002}, the generated Raman comb usually has a complex spectrum shape and is unstable. Here we studied the detailed dynamics of Raman comb formation in silica rod microresonators, which have cavity free spectral ranges (FSRs) in microwave rates. We focused particularly on the frequency shift within the Raman gain with two large peaks at 13.2~THz (Peak~1) and 14.7~THz (Peak~2). Silica microresonators with a small cavity FSR (large diameter) were used for this study because a large number of resonant modes are present within the Raman gain spectrum. In our experiment, we excited one of the specific peaks and tried to generate a Raman comb with a smooth spectral envelope by controlling the pump detuning and the coupling strength. The controllability of a Raman comb is the key to various applications. In addition, a good understanding of the Raman comb formation is also important for comb generation via FWM because these processes compete inside a microresonator \cite{Min_APL_2005,Chembo_PRA_2015,Milian_PRA_2015,Karpov_PRL_2016,Yang_NP_2017,Okawachi_OL_2017,Fujii_OE_2017}.\par
The paper is organized as follows. The experiment is described in \S \ref{sec:2}, where the generation of a smooth Raman comb and the peak transition are studied. Then a numerical approach based on coupled mode equations is used to explain the peak transition behavior in \S \ref{sec:3}. Finally, we conclude in \S \ref{sec:4}.

\section{Experiment} \label{sec:2}

\begin{figure}[t]
	\centering
	\fbox{\includegraphics[width=85mm]{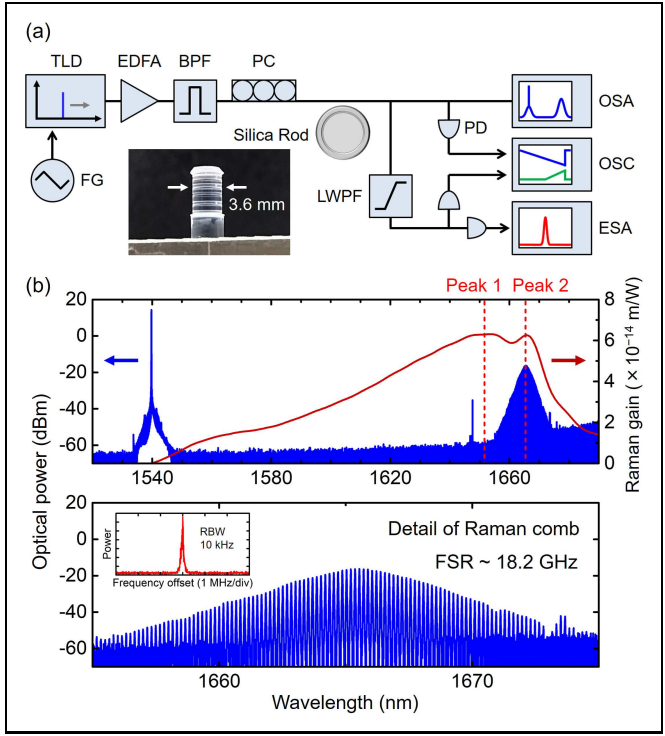}}
	\caption{(a)~Experimental setup for Raman comb generation. TLD: tunable laser diode, FG: electrical function generator, EDFA: erbium doped fiber amplifier, BPF: band pass filter, PC: polarization controller, LWPF: long wave pass filter, PD: photodetector, OSA: optical spectrum analyzer, OSC: oscilloscope, ESA: electrical spectrum analyzer. The inset shows a silica rod microresonator with a cavity FSR of 18.2~GHz. (b)~Optical spectra (blue) and a theoretical Raman gain pumped at 1540~nm (red). Raman gain has two large peaks with frequency shifts of 13.2~THz (Peak~1) and 14.7~THz (Peak~2). The pump frequency does not affect the amount of frequency shift but affects the gain intensity (proportional to the pump frequency). The generated Raman comb had a smooth spectral envelope whose center wavelength corresponded to Peak~2. The inset shows the beatnote signal measured by detecting the generated Raman comb. The 3~dB linewidth is less than hundreds kHz.}
	\label{fig:1}
\end{figure}

\begin{figure}[t]
	\centering
	\fbox{\includegraphics[width=82mm]{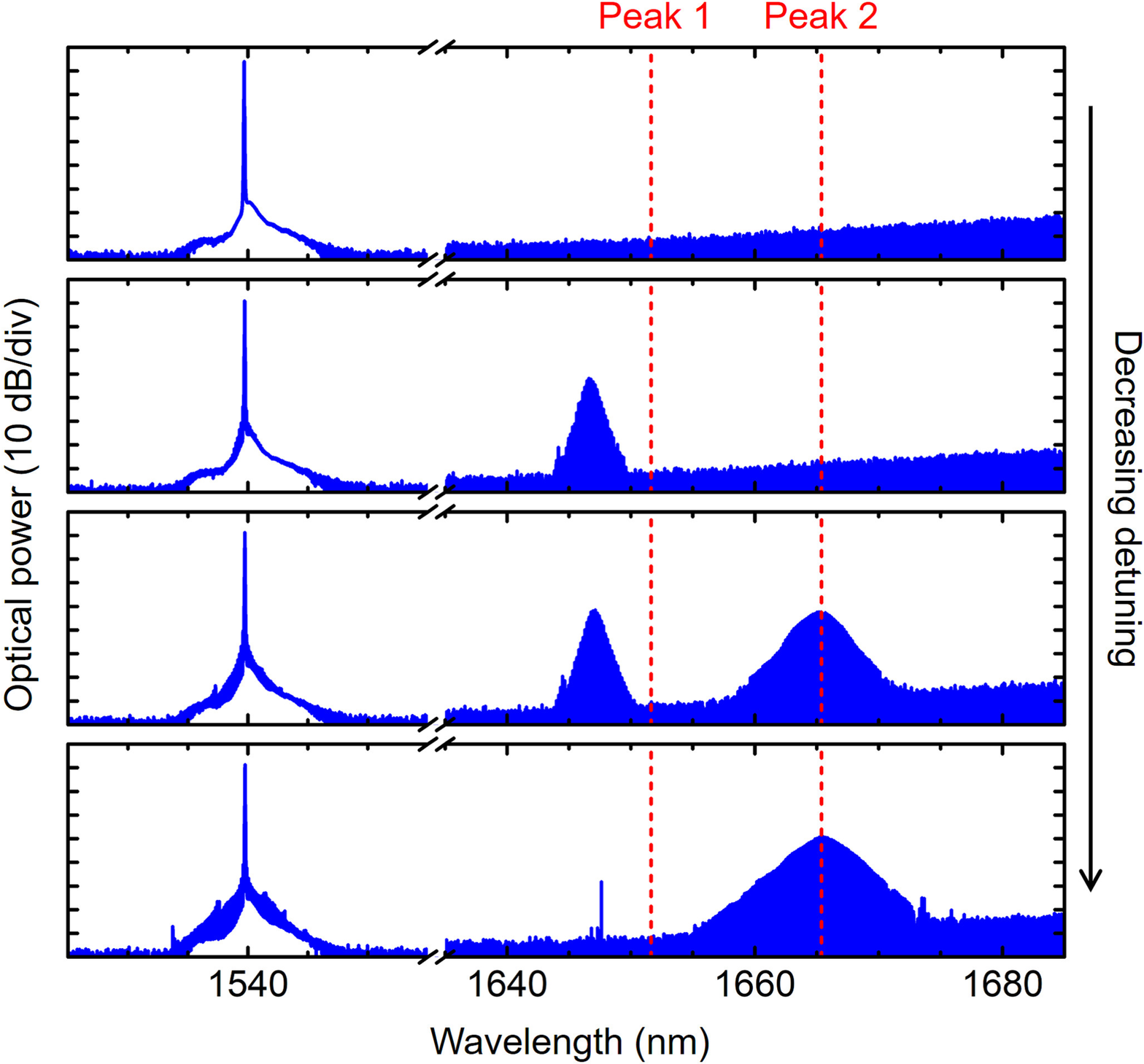}}
	\caption{Raman comb formation while decreasing the detuning between pump and resonance frequencies. The Raman offset transited from Peak~1 to Peak~2. The pump wavelength and input power were 1540~nm and 80~mW, respectively.}
	\label{fig:2}
\end{figure}

Figure~\ref{fig:1}(a) shows our experimental setup. A CW laser light with 100~kHz linewidth was amplified and then coupled to a silica rod microresonator by using a tapered fiber. The coupling strength was adjusted by changing the fiber position. Silica rod microresonators with an intrinsic Q ($Q_{\rm i}$) over 100 million were fabricated by using CO$_2$ laser processes \cite{DelHaye_APL_2013,Papp_PRX_2013}. We used a microresonator with a diameter of 3.6~mm, which corresponds to a cavity FSR of 18.2~GHz. The pump light was scanned from a short to a long wavelength to generate a Raman comb, because the resonance was shifted via thermal effects \cite{Carmon_OE_2004}. The output light was monitored with an optical spectrum analyzer. In addition, the Raman comb was employed to measure its beatnote signal by using a 20~GHz photodetector connected to an electrical spectrum analyzer. Figure~\ref{fig:1}(b) shows the measured optical spectra obtained while generating a Raman comb with an under-coupling condition. The pump energy at 1540~nm was transferred to components with a longer wavelength of around 1665~nm via SRS. The red line shows the theoretical Raman gain spectrum when the pump is located at 1540~nm. With 1540.0~nm pumping, the two peaks, Peaks~1 (13.2~THz shift) and 2 (14.7~THz shift), are located at wavelengths of 1651.9 and 1665.7~nm, respectively. Due to the complex shape of the Raman gain spectrum, a Raman comb with an asymmetric spectrum is usually excited when parameters such as the excitation power and the coupling condition between the microresonator and fiber are not adjusted. Only when we carefully adjust the parameters, do we obtain a Raman comb with a smooth spectral envelope as shown in Fig.~\ref{fig:1}(b). In this experiment, higher order cascaded SRS was hardly observed.\par
First, we investigated the transition of the Raman comb position between two peaks, Peaks~1 and 2, within the Raman gain spectrum. Figure~\ref{fig:2} shows spectrum formation via SRS while decreasing the detuning between the resonance and pump frequencies. The pump light with large detuning generated SRS close to the Peak~1 wavelength. Decreasing the detuning caused SRS in a longer wavelength regime, whose offset from the pump matched Peak~2. By further decreasing the detuning, we observed an energy transfer from Peak~1 to Peak~2 and the suppression of the Raman light at Peak~1. The Raman energy transition was also observed with another silica rod microresonator, which has a cavity FSR of 32.4~GHz. Hence, by controlling the detuning, we can selectively generate a Raman comb with an offset wavelength at Peak~1 or Peak~2. Such behavior is also observed in silica fibers when the input power is changed \cite{Stolen_JOSAB_1984}. Although a detailed explanation is given in \S \ref{sec:3}, it can be described briefly as follows. Since the gain of Peak~1 is slightly higher than that of Peak~2, Peak~1 reaches the SRS threshold first. As Peak~2 is excited directly from the pump and light generated at Peak~1, it reaches the lasing threshold as the pump power is further increased.\par
With a silica microresonator with an anomalous dispersion, the competition between the SRS and FWM thresholds is given as $P_{\rm in}^{\rm SRS}/P_{\rm in}^{\rm FWM}\approx(2\omega_0 n_2)/(c g_{\rm R(max)})\approx2.9$, which implies that the FWM occurs before generating SRS \cite{Spillane_N_2002,Herr_NP_2012}. $\omega_0$ is the pump mode frequency, $n_2$ is the nonlinear refractive index of silica, $c$ is the speed of light, and $g_{\rm R(max)}$ is the maximum Raman gain of silica in m/W units. However, in our experiment, typical degenerate FWM caused by modulation instability did not occur while changing the detuning. Although the mode family used in this experiment has an anomalous dispersion, the mode coupling changes it so that it exhibits an effective normal dispersion locally around the pump. Hence mode couplings with other mode families disturbed FWM generation. The mode family property was confirmed experimentally by performing a dispersion measurement with a fiber Mach-Zehnder interferometer calibration \cite{Huang_PRL_2015,Suzuki_OE_2017}.\par
\begin{figure}[t]
	\centering
	\fbox{\includegraphics[width=73mm]{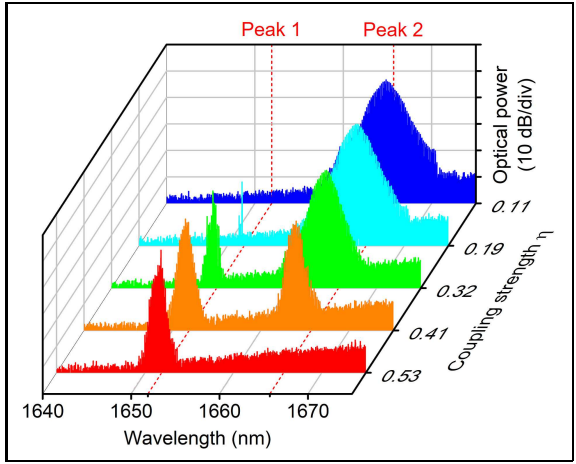}}
	\caption{Raman comb spectra depending on the coupling strength (fiber position) with near zero pump detuning. A weaker (stronger) coupling condition induced the generation of a Raman comb at Peak~2 (Peak~1). The pump wavelength and input power were 1540~nm and 160~mW, respectively.}
	\label{fig:3}
\end{figure}

Next, we investigated coupling strength dependency. We controlled the coupling strength by changing the tapered fiber position and measured the comb spectra with near zero pump detuning (Fig.~\ref{fig:3}). The coupling strength is defined as $\eta=\kappa_{\rm c}/\kappa$ ($\eta=0.5$ denotes critical coupling). $\kappa$ and $\kappa_{\rm c}$ are the cavity decay rate and the coupling rate with the waveguide, respectively. $\kappa$ has the relation as $Q=\omega_0/\kappa$ and $\kappa=\kappa_{\rm i}+\kappa_{\rm c}$. $Q$ is the effective cavity Q and $\kappa_{\rm i}$ is the intrinsic loss rate. Figure~\ref{fig:3} shows that weak coupling induced the generation of a Raman comb at Peak~2, while strong coupling results in the excitation of Peak~1. The coupling strength values are obtained from the linewidth of the measured cavity resonance modes.\par

\begin{figure}[t]
	\centering
	\fbox{\includegraphics[width=82mm]{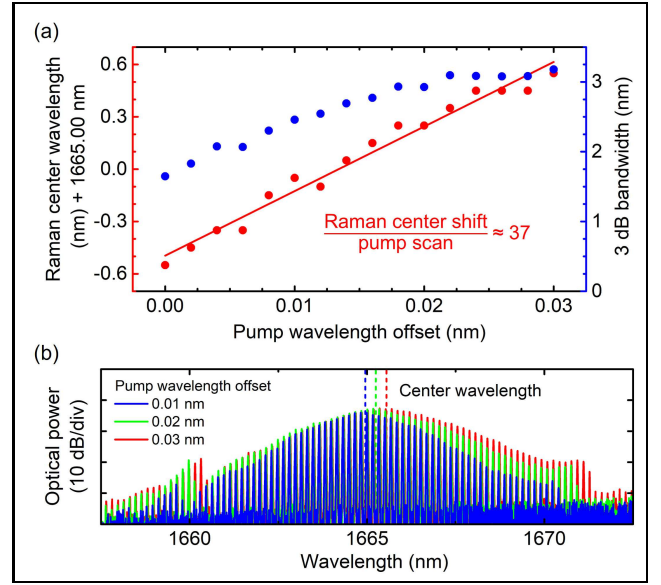}}
	\caption{(a)~The center wavelength (red) and 3~dB bandwidth (blue) in a Raman comb generated at Peak~2 as a function of the pump wavelength. (b)~Raman comb spectra with different pump wavelengths (detuning). The dashed lines represent the center wavelength of each comb envelope.}
	\label{fig:4}
\end{figure}

Longer wavelength modes are excited from modes at shorter wavelengths, that induces the center wavelength of a Raman comb shifts via SRS. Figure~\ref{fig:4}(a) shows the center wavelength (red) and 3~dB bandwidth (blue) of a Raman comb as a function of the pump wavelength. The amount of the center wavelength shift was much larger than that of pump scanning, whose ratio is about 37. Figure~\ref{fig:4}(b) shows Raman comb spectra with different pump wavelengths, which indicate that the comb envelope broadens to a longer wavelength. The center wavelength shift of a mode-locked soliton pulse (corresponding to the pulse delay in the time domain) has been observed in optical fibers \cite{Skryabin_RMP_2010} and microresonators \cite{Karpov_PRL_2016}. Since the Raman comb generated in this experiment did not propagate as a soliton pulse, the ratio of the center wavelength shift was small.\par
To examine the coherent property of the output light, we measured the beatnote signals with a fast photodetector. It can be directly measured because the generated Raman comb has mode spacings in microwave rates. The measured beatnote signals exhibit multiple peaks when the output is not a smooth spectrum, because Raman light is generated in various mode families. On the other hand, when we obtain a Raman comb with a smooth spectrum envelope, it has only one peak with a 3~dB linewidth of less than hundreds kHz as shown in the inset of Fig.~\ref{fig:1}(b). The linewidth is limited by the presence of the dispersion, because each SRS line is located at the center of each resonance mode. Although a linewidth in the kHz range cannot be regarded as phase locking, it has the potential to obtain smooth and phase-locked Raman combs through optimizing the operation wavelength, cavity dispersion, and stabilization techniques. In particular, according to previous reports \cite{Liang_PRL_2010,Chembo_PRA_2015}, pumping in a weak normal dispersion regime is suitable for generating phase-locked Raman combs.

\section{Coupled mode equations with Raman processes} \label{sec:3}
To study the energy exchange between the pump and other resonant modes covered by Raman gain, we calculated the numbers of intracavity photons by using coupled mode equations taking Raman processes into account. This calculation is used to obtain the dynamics of energy exchange via SRS, as functions of coupling strength and detuning between pump and resonant frequencies. The internal field in each mode $a_m$ ($m=0,1,2,\cdots$) has the resonance frequency $\omega_m=\omega_0-mD_1$, where $\omega_0$ is the pump mode frequency and $D_1$ is the cavity FSR in rad$\cdot$Hz units. The time evolution of these fields is given by
\begin{equation} 	\label{eq:1}
\begin{split}
\frac{\partial a_m}{\partial t}=&-\frac{1}{2}\kappa a_m+\sum_{m',0{\leq}m'<m}G_{{\rm R}(m',m)}|a_{m'}|^2a_m\\
&-\sum_{m',m<m'}G_{{\rm R}(m,m')}|a_{m'}|^2a_m~\Bigl[+i\Delta\omega_0 a_m+\sqrt{\kappa_{\rm c}}s_{\rm in}\Bigr]_{m=0}.
\end{split}
\end{equation}
Here $s_{\rm in}$ is the input field $|s_{\rm in}|^2=P_{\rm in}/(\hbar\omega_0)$, $P_{\rm in}$ is the power input into the coupling waveguide, $\hbar$ is the Planck constant divided by 2$\pi$, and $\Delta\omega_0$ represents the pump detuning from the resonance. The input field is only added to the equation for pump mode ($m=0$). The number of intracavity photons in each mode corresponds to $|a_m|^2$, which can be converted to power by $\hbar\omega_m|a_m|^2\times D_1/2\pi$. The nonlinear Raman coefficient is written as
\begin{equation}
G_{{\rm R}(j,k)}=g_{\rm R}(\omega_j,\omega_k)\frac{c D_1\hbar\omega_j}{4\pi n A_{\rm eff}},
\end{equation}
where $j$ and $k$ represent the mode numbers ($j,k\in0,1,2,\cdots$), $n$ is the effective refractive index of the resonance mode, and $A_{\rm eff}$ is the effective nonlinear mode area. $g_{\rm R}(\omega_j,\omega_k)$ represents the Raman gain in silica with the frequency shift $|\omega_k-\omega_j|$ when pumping at a frequency $\omega_j$. For 1550~nm pumping with silica, the Raman gain value is $6.2\times10^{-14}$~m/W at a 13.2~THz shift \cite{Agrawal_2007,Hollenbeck_JOSAB_2002}. The gain value is in proportional to the pump frequency. $\kappa$, $\kappa_{\rm c}$, $\kappa_{\rm i}$, and $A_{\rm eff}$ are regarded as independent of the mode number $m$. \par

\begin{table}[t]
	\centering
	\caption{Parameters used for the calculation}
	\begin{tabular}{cccccc} \hline
		$\lambda_0$  & $P_{\rm in}$ &$D_1/2\pi$ & $Q_{\rm i}$ & $n$ & $A_{\rm eff}$ \\ \hline
		1540~nm & 160~mW & 18.2~GHz & 1$\times10^8$ & 1.44 & 135~\textmu m$^2$  \\ \hline
	\end{tabular}
	\label{table:1}
\end{table}

\begin{figure}[t]
	\centering
	\fbox{\includegraphics[width=85mm]{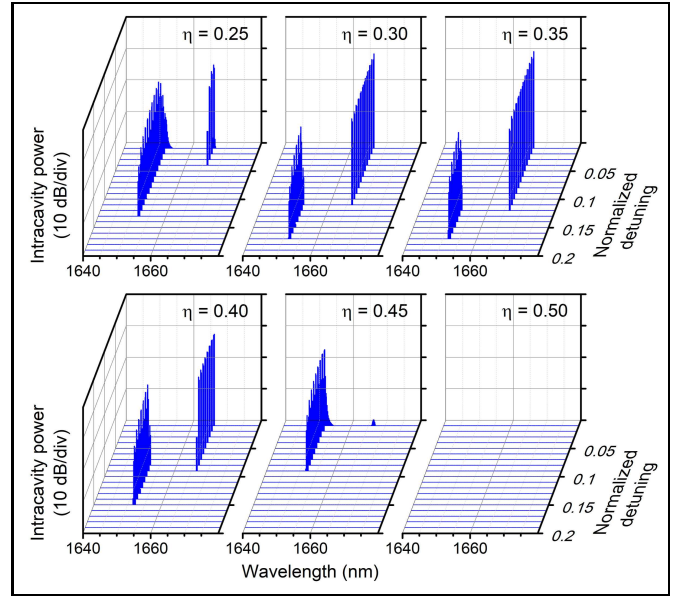}}
	\caption{Transition of SRS wavelength, calculated by using Eq.~(\ref{eq:1}). The pump wavelength is 1540.0~nm (not shown in this figure). Peaks~1 and 2 of the Raman gain correspond to 1651.9 and 1665.7~nm, respectively. A clear SRS wavelength transition is observed at a coupling strength ($\eta$) of 0.30-0.40. On the other hand, SRS does not occur due to the relatively low effective Q at a critical coupling condition ($\eta=0.5$) with the parameters shown in Table~\ref{table:1}.}
	\label{fig:5}
\end{figure}

Figure~\ref{fig:5} shows the lasing action for different coupling conditions as a function of the detuning, which is normalized by $\kappa$. The calculation parameters are shown in Table~\ref{table:1}, which follow our experimental condition. $A_{\rm eff}$ is calculated by using a mode solver based on the finite element method. We set the number of modes at 1001, which corresponds to a wavelength range of 1540.0 to 1698.8~nm. With larger detuning, SRS does not occur because the intracavity power is below the threshold. After exceeding the threshold, red-shifted light starts to appear close to Peak~1 in the Raman gain (1651.9~nm). This is because the gain value at Peak~1 is larger than that at Peak~2. When the detuning is even smaller, the SRS wavelength moves close to the Peak~2 (1665.7~nm) due to the simultaneous excitation by two components; i.e. by the pump and Raman light at Peak~1. The calculation shows the clear transition of the SRS light from one wavelength to another, which is dependent on the detuning.\par
In addition, the peak transition is dependent on the coupling strength. When the coupling strength is weak, SRS is generated at Peak~2. When the coupling strength is even weaker, the system does not exhibit SRS because the intracavity pump power does not reach the threshold. On the other hand, the critical coupling condition does not allow us to achieve SRS due to the relatively low effective Q. The calculation results show that a smaller detuning and weaker coupling strength are needed to generate SRS efficiently, and the result agrees with our experimental observations shown in Figs.~\ref{fig:2} and \ref{fig:3}. Although weak FWM comb lines (i.e. close to the pump light) were observed in our experiment as shown in Fig.~\ref{fig:2}, they do not greatly influence the result because the unequal spacings of the SRS comb lines do not strongly interact via FWM. Therefore, our model where no FWM processes are taken into account is sufficient to describe the phenomena.\par
We observed some differences between the experiment and the numerical analysis. We observed multi-mode lasing in the experiment, while the system exhibited lasing at a few modes in our numerical analysis. This is because the numerical analysis considers only one mode as a pump, and that mode is modeled with the mean field in a microresonator. The pump intensity is clamped after the SRS occurs and this inhibits the lasing of other resonant modes, because all SRS modes share the same pump mode. However, in the experiment, all the molecules at different sites exhibit different vibration frequencies, which is the origin of the inhomogeneous broadening of the Raman gain \cite{Hollenbeck_JOSAB_2002}. Hence, we obtained multi-mode Raman lasing in the experiment. Despite this difference, the energy transition behavior from Peak~1 to Peak~2 is explained by using our model.

\section{Conclusion} \label{sec:4}
In conclusion, we generated Raman combs from a silica rod microresonator with an 18.2~GHz FSR and controlled the center wavelength via detuning and coupling optimization. In the experiment, reducing the detuning leads to the Raman energy transition from Peak~1 to Peak~2, which is similar behavior to that observed in silica fibers. Also, weak coupling induced to generate a Raman comb at Peak~2, while strong coupling results in only the excitation of Peak~1. Those observations are in agreement with those of a simulation, which were calculated by using coupled mode equations taking Raman processes into account. In addition, we observed the center wavelength shift of a Raman comb, with a shift that is 37 times larger than that of pump scanning.\par
This study provides an explanation of the formation dynamics, which must be understood if we are to generate smooth and phase-locked Raman combs. Since phase-locked Raman combs can be generated in a normal dispersion regime, it will expand phase-locked comb generation towards the visible wavelength regime, which is difficult to demonstrate with conventional FWM processes. Such phase-locked Raman combs can be used for applications such as compact pulse laser sources, sensors, optical clocks, and coherence tomography.

\section*{Funding.}
This work was supported by JSPS KAKENHI Grant Numbers JP16J04286, JP15H05429.

\section*{Acknowledgment.}
The authors would like to thank Dr. T.~Kato for valuable discussion. First author acknowledges the Program for Leading Graduate Schools, ``Global Environmental System Leaders Program'' by the Ministry of Education, Culture, Sports, Science and Technology (MEXT) in Japan.


\begin{thebibliography}{99}
	\newcommand{\enquote}[1]{``#1''}
	
	\bibitem{Stolen_APL_1977}
	R.~H. Stolen, C.~Lin, and R.~K. Jain, \enquote{A time-dispersion-tuned
		fiber Raman oscillator,} Appl. Phys. Lett. \textbf{30}, 340--342 (1977).
	
	\bibitem{Stolen_JOSAB_1984}
	R.~H. Stolen, C.~Lee, and R.~K. Jain, \enquote{Development of the stimulated
		Raman spectrum in single-mode silica fibers,} J. Opt. Soc. Am. B \textbf{1},
	652--657 (1984).
	
	\bibitem{Boyraz_OE_2004}
	O.~Boyraz and B.~Jalali, \enquote{Demonstration of a silicon Raman laser,} Opt.
	Express \textbf{12}, 5269--5273 (2004).
	
	\bibitem{Spillane_N_2002}
	S.~M. Spillane, T.~J. Kippenberg, and K.~J. Vahala, \enquote{Ultralow-threshold
		Raman laser using a spherical dielectric microcavity,} Nature \textbf{415},
	621--623 (2002).
		
	\bibitem{Kippenberg_IEEEJSTQE_2004}
	T.~J.~Kippenberg, S.~M.~Spillane, B.~Min, and K.~J.~Vahala, \enquote{Theoretical and experimental study of stimulated and cascaded Raman scattering in ultrahigh-Q optical microcavities,} IEEE J. Sel. Top. Quantum Electron. \textbf{10}, 1219--1228 (2004). 
	
	\bibitem{Ozdemir_PNAS_2014}
	{\c{S}}.~K.~{\"O}zdemir, J.~Zhu, X.~Yang, B.~Peng, H.~Yilmaz, L.~He, F.~Monifi, S.~H.~Huang, G.~L.~Long, and L.~Yang, \enquote{Highly sensitive detection of nanoparticles with a self-referenced and self-heterodyned whispering-gallery Raman microlaser,} P. Natl. Acad. Sci. USA \textbf{111}, E3836--E3844 (2014).
	
	\bibitem{Ooka_APE_2015}
	Y.~Ooka, Y.~Yang, J.~Ward, and S.~N. Chormaic, \enquote{Raman lasing in a
		hollow, bottle-like microresonator,} Appl. Phys. Express \textbf{8}, 092001
	(2015).
	
	\bibitem{Kato_OE_2017}
	T.~Kato, A.~Hori, R.~Suzuki, S.~Fujii, T.~Kobatake, and T.~Tanabe,
	\enquote{Transverse mode interaction via stimulated Raman scattering comb in
		a silica microcavity,} Opt. Express \textbf{25}, 857--866 (2017).
	
	\bibitem{Yang_NP_2017}
	Q.-F. Yang, X.~Yi, K.~Y. Yang, and K.~Vahala, \enquote{Stokes solitons in
		optical microcavities,} Nat. Phys. \textbf{13}, 53--57 (2017).
	
	\bibitem{Rong_NP_2007}
	H.~Rong, S.~Xu, Y.-H. Kuo, V.~Sih, O.~Cohen, O.~Raday, and M.~Paniccia,
	\enquote{Low-threshold continuous-wave Raman silicon laser,} Nat. Photonics
	\textbf{1}, 232--237 (2007).
	
	\bibitem{Rong_NP_2008}
	H.~Rong, S.~Xu, O.~Cohen, O.~Raday, M.~Lee, V.~Sih, and M.~Paniccia, 
	\enquote{A cascaded silicon Raman laser,} Nat. Photonics \textbf{2}, 170--174 (2008).
	
	\bibitem{Grudinin_OL_2007}
	I.~S. Grudinin and L.~Maleki, \enquote{Ultralow-threshold Raman lasing with
		CaF$_2$ resonators,} Opt. Lett. \textbf{32}, 166--168 (2007).
	
	\bibitem{Liang_PRL_2010}
	W.~Liang, V.~S. Ilchenko, A.~A. Savchenkov, A.~B. Matsko, D.~Seidel, and
	L.~Maleki, \enquote{Passively mode-locked Raman laser,} Phys. Rev. Lett.
	\textbf{105}, 143903 (2010).
	
	\bibitem{Lin_OL_2016}
	G.~Lin and Y.~K. Chembo, \enquote{Phase-locking transition in Raman combs
	generated with whispering gallery mode resonators,} Opt. Lett. \textbf{41},
	3718--3721 (2016).
		
	\bibitem{Vanier_OE_2014}
	F.~Vanier, Y.~A.~Peter, and M.~Rochette, \enquote{Cascaded Raman lasing in packaged high quality As$_2$S$_3$ microspheres,} Opt. Express  \textbf{22}, 28731--28739 (2014). 
	
	\bibitem{Li_OL_2013}
	B.-B. Li, Y.-F. Xiao, M.-Y. Yan, W.~R. Clements, and Q.~Gong,
	\enquote{Low-threshold Raman laser from an on-chip, high-Q, polymer-coated
		microcavity,} Opt. Lett. \textbf{38}, 1802--1804 (2013).
	
	\bibitem{Latawiec_O_2015}
	P.~Latawiec, V.~Venkataraman, M.~J. Burek, B.~J. Hausmann, I.~Bulu, and
	M.~Lon{\v{c}}ar, \enquote{On-chip diamond Raman laser,} Optica \textbf{2},
	924--928 (2015).
	
	\bibitem{Liu_O_2017}
	X.~Liu, C.~Sun, B.~Xiong, L.~Wang, J.~Wang, Y.~Han, Z.~Hao, H.~Li, Y.~Luo, J.~Yan, T.~Wei, Y.~Zhang, and J.~Wang, \enquote{Integrated continuous-wave aluminum nitride Raman laser,} Optica \textbf{4}, 893--896 (2017). 
	
	\bibitem{Agrawal_2007}
	G.~P.~Agrawal, \textit{Nonlinear fiber optics}  (Academic, 2007). 
	
	\bibitem{Hollenbeck_JOSAB_2002}
	D.~Hollenbeck and C.~D. Cantrell, \enquote{Multiple-vibrational-mode model for
		fiber-optic Raman gain spectrum and response function,} J. Opt. Soc. Am. B
	\textbf{19}, 2886--2892 (2002).
	
			\bibitem{Min_APL_2005}
	B.~Min, L.~Yang, and K.~Vahala, \enquote{Controlled transition between parametric and Raman oscillations in ultrahigh-Q silica toroidal microcavities,} Appl. Phys. Lett. \textbf{87}, 181109 (2005).
	
	\bibitem{Chembo_PRA_2015}
	Y.~K.~Chembo, I.~S.~Grudinin, and N.~Yu, \enquote{Spatiotemporal dynamics of Kerr-Raman optical frequency combs,} Phys. Rev. A \textbf{92}, 043818 (2015).
	
	\bibitem{Milian_PRA_2015}
	C.~Mili\'an, A.~V.~Gorbach, M.~Taki, A.~V.~Yulin, and D.~V.~Skryabin, \enquote{Solitons and frequency combs in silica microring resonators: Interplay of the Raman and higher-order dispersion effects,} Phys. Rev. A \textbf{92}, 033851 (2015).
	
	\bibitem{Karpov_PRL_2016}
	M.~Karpov, H.~Guo, A.~Kordts, V.~Brasch, M.~H.~P. Pfeiffer, M.~Zervas,
	M.~Geiselmann, and T.~J. Kippenberg, \enquote{Raman self-frequency shift of
		dissipative Kerr solitons in an optical microresonator,} Phys. Rev. Lett.
	\textbf{116}, 103902 (2016).
	
	\bibitem{Okawachi_OL_2017}
	Y.~Okawachi, M.~Yu, V.~Venkataraman, P.~M.~Latawiec, A.~G.~Griffith, M.~Lipson, M.~Lon{\v{c}}ar, and A.~L.~Gaeta, \enquote{Competition between Raman and Kerr effects in microresonator comb generation,} Opt. Lett. \textbf{42}, 2086 (2017).

	\bibitem{Fujii_OE_2017}
	S.~Fujii, T.~Kato, R.~Suzuki, A.~Hori, and T.~Tanabe, \enquote{Transition between Kerr comb and stimulated Raman comb in a silica whispering gallery mode microcavity,} J. Opt. Soc. Am. B \textbf{35}, 100--106 (2018).

	\bibitem{DelHaye_APL_2013}
	P.~Del'Haye, S.~A. Diddams, and S.~B. Papp, \enquote{Laser-machined
		ultra-high-Q microrod resonators for nonlinear optics,} Appl. Phys. Lett.
	\textbf{102}, 221119 (2013).
	
	\bibitem{Papp_PRX_2013}
	S.~B. Papp, P.~Del'Haye, and S.~A. Diddams, \enquote{Mechanical control of a
		microrod-resonator optical frequency comb,} Phys. Rev. X \textbf{3}, 031003
	(2013).
	
	\bibitem{Carmon_OE_2004}
	T.~Carmon, L.~Yang, and K.~J.~Vahala, \enquote{Dynamical thermal behavior and thermal self-stability of microcavities,}
	 Opt. Express \textbf{12}, 4742--4750 (2004).
	
	\bibitem{Herr_NP_2012}
	T.~Herr, K.~Hartinger, J.~Riemensberger, C.~Y.~Wang, E.~Gavartin, R.~Holzwarth, M.~L.~Gorodetsky, and T.~J.~Kippenberg, \enquote{Universal formation dynamics and noise of Kerr-frequency combs in microresonators,} Nat. Photonics \textbf{6}, 480--487 (2012).
		
	\bibitem{Suzuki_OE_2017}
	R.~Suzuki, T.~Kato, T.~Kobatake, and T.~Tanabe, \enquote{Suppression of optomechanical parametric oscillation in a toroid microcavity assisted by a Kerr comb,} Opt. Express \textbf{25}, 28806--28816 (2017).
	
	\bibitem{Huang_PRL_2015}
	S.-W.~Huang, H.~Zhou, J.~Yang, J.~F.~McMillan, A.~Matsko, M.~Yu, D.-L.~Kwong, L.~Maleki, and C.~W.~Wong, \enquote{Mode-locked ultrashort pulse generation from on-chip normal dispersion microresonators,} Phys. Rev. Lett. \textbf{114}, 053901 (2015).
		
	\bibitem{Skryabin_RMP_2010}		
		D.~V.~Skryabin and A.~V.~Gorbach, \enquote{Colloquium: Looking at a soliton through the prism of optical supercontinuum,} Rev. Mod. Phys. \textbf{82}, 1287--1299 (2010).

	

\end{thebibliography}
\end{document}